# Tilting behavior of lamellar ice tip during unidirectional freezing of aqueous solutions

*Tongxin Zhang, Zhijun Wang\*, Lilin Wang\*, Junjie Li, Jincheng Wang*

*State Key Laboratory of Solidification Processing, Northwestern Polytechnical University, Xi'an 710072, China*

**Abstract:** Freezing of ice has been largely reported from many aspects, especially its complex pattern formation. Ice grown from liquid phase is usually characteristic of lamellar morphology which plays a significant role in various domains. However, tilted growth of ice via transition from coplanar to non-coplanar growth in directional solidification has been paid little attention in previous studies and there is misleading explanation of the formation of tilted lamellar ice. Here, we in-situ investigated the variations of tilting behavior of lamellar ice tip under different conditions within a single ice crystal with manipulated orientation via unidirectional freezing of aqueous solutions. It is found that tilted growth of ice tips is sensitive to pulling velocity and solute type. These experimental results reveal intrinsic tilted growth behavior of lamellar ice and enrich our understanding in pattern formation of ice.

## 1. Introduction

Pattern formation during ice growth has been widely studied in the past decades among a great variety of both scientific and engineering communities due to its significance in both condensed matter physics [1, 2], atmospheric science [3], icephobic materials design [4, 5], material processing engineering [6-8] and cryobiology [9-12]. Although there are plenty of advances in the past centuries, one interesting but puzzling growth habit of ice is the occurrence of tilted growth of ice cell/dendrite in undercooled pure water [13] and aqueous solutions [14-16] distinct from a planar dendrite in its basal plane at small undercooling. In both free growth and unidirectional growth experiments, lamellar ice in the form of either an array of dendrites [17] or a single ice dendrite [18] was usually reported to grow in a "noncrystallographic" [19] or, in other words, a tilted direction deviating from basal plane. To be more specific, ice grows as a planar disk at small undercoolings but grows in a non-coplanar manner with a tilt angle deviating from its basal plane {0001} at larger bath undercooling of about 2.5 K [13, 18]. Furthermore, solute additives were also reported to quantitatively alter the dependence of tilt angle on undercooling and growth velocity. Thus, tilted growth of ice has an intrinsically different nature compared to traditional tilted growth of metals and alloys [20, 21] in metallurgical investigations which deals with the resulting morphology with unparalleled thermal gradient and preferred crystal orientation.

---

\* Corresponding author. zhjwang@nwpu.edu.cn
\* Corresponding author. wlilin@nwpu.edu.cn

Tilt angle is a direct indicator of tilted growth of an ice dendrite with respect to its basal pane. Roughly speaking, previous investigations on tilt angle of ice can be divided into two categories. One is ice growth kinetics data of basal and edge planes of ice in various growth conditions, and it is reasonable to predict tilt angle of ice based on these ice growth data in combination with a proposed "step growth mechanism"[13]. The other intuitively belongs to direct measurement of tilt angle within a single ice crystal mostly from free growth experiments. One of the problems in these reports is the complexity of growth kinetics laws for basal and edge planes. In most cases, researchers will assume a lateral growth mechanism for basal plane and a continuous growth mechanism for edge plane due to their interface roughness. Some others further found a change of growth kinetics of basal plane when undercooling exceeds certain values [22]. And a few authors [16, 23] even noticed that a transition of growth kinetics for edge plane should also exist. The other problem is the scattered results of tilt angle which are not strictly or even hardly comparable among different researchers. However, the consensus on tilt angle of ice is that, within a limited range of bath undercooling investigated, tilt angle will first increase and reach a plateau or a limiting value which will no longer increase with undercooling. Although the maximum tilt angle $\alpha$ in many free growth experiments is 22.5°, it was reported that $\alpha$ can reach a limiting value of as large as 38° or 45° and the anisotropic growth in basal and edge plane would disappear, according to numerous wind tunnel experiments by Levi [24] and an exceptional case of water drop freezing experiments by Hallett [25]. More systematic investigations in combination with crystal growth theory are needed to enrich our understanding of ice growth kinetics in basal and edge planes.

In contrast to free growth investigations of ice, unidirectional growth of ice especially within a single ice crystal has received far less attention despite its significance in ice morphology. In previous investigations [26], tilting of actual growth direction of ice tip can occur even under an initially parallel thermal gradient with respect to its basal plane, which validates a tilting behavior different from the combined effect of unparalleled thermal gradient and preferred orientation of ice. There is no doubt that unidirectional growth of ice in relation with its tilted growth is worth further investigation in hope of gaining a deeper understanding of pattern formation during unidirectional ice growth. Thus, in this paper we look for more details on how this tilting behavior of ice responds to unidirectional growth parameters such as thermal gradient, growth velocity, solute concentration and solute type. In this paper, in-situ unidirectional experiments in combination with controlled ice orientation are carried out.

## 2. Experimental setup

NaCl solutions (0.1M, 0.2M, 0.3M and 0.6M) and PVA103/203 solutions (5 wt.% and 10wt.%) were prepared in the method reported elsewhere [27]. Rectangular glass capillary with an inner space dimension of $0.05 \times 1$ mm$^2$ (VitroCom brand) was utilized for unidirectional freezing experiments. More details on the experimental setup can be found in Ref. [26]. Different thermal gradients were produced by changing the temperature on cold and hot sides of unidirectional freezing platform. Prior to each freezing experiments, single ice crystal was prepared with an orientation parallel to the direction of thermal gradient and pulling velocity of the step motor.

We obtained the solid/liquid (S/L) interface morphology and measured the ice tip position and interface/tip undercooling during steady state growth in which the S/L interface position remained unchanged in our in-situ experiments. The time to reach steady state growth of lamellar ice usually takes a few hundred seconds in our experiments. And when steady state growth is reached, the growth velocity of S/L interface equals pulling velocity. The measured tip position as a function of time is used to precisely describe the tilted angle of the ice tip. The effect of varying thermal gradient is first investigated for 0.1 M NaCl solution with increasing pulling velocity. And experiments are latter carried out in NaCl solutions with different concentrations (0.1M, 0.2M, 0.3M and 0.6M) and also in PVA103/203 solutions with different concentrations (5 wt.% and 10wt.%). In addition, we recorded morphology evolution of ice after an abrupt stop from steady state freezing in 0.2 M NaCl solution and 10 wt.% PVA203 solution to examine the evolution of tilted growth morphology after different time of intervals upon stop.

## 3. Results and discussions

### *3.1 Typical morphologies of S/L interface*

To obtain an intuitive sense of the experimental results, we have summarized tilted growth morphologies in NaCl solutions in **Fig. 1**. The figures **(a1)**, **(b1)** and **(c1)** represent the different experiments with slightly different ice orientations, where **(a2)**, **(b2)** and **(c2)** are these with larger growth velocities compared with **(a1)**, **(b1)** and **(c1)**. A short experimental movie for growth conditions in **Fig. 1 (a2)** is provided in **Appendix A**. In **Fig. 1 (a1-a2)**, the basal plane {0001} of single ice crystal (represented by red dashed arrow) was nearly parallel to pulling velocity ($V_p$) / thermal gradient ($G$) (represented by black solid arrow on top of **Fig. 1**) and also parallel to incident light ($\vec{L}$), whereas $V_p(G)$ is perpendicular to $\vec{L}$, giving the relationships of $\vec{V}_P \parallel \vec{G} \parallel \{0001\}, \vec{L} \parallel \{0001\}, \vec{V}_p(\vec{G}) \perp \vec{L}$. Tilting of lamellar ice arrays produces two groups of ice tips with different shapes of ice tip. Tilting directions of the two groups of ice tips (represented by two red solid arrows in each figure of **Fig. 1**) with respect to $V_p/G$ is symmetrical (tilting up or down). In **Fig. 1 (a2)**, when pulling velocity increases, tilt angle $\alpha$ increases correspondingly. In **Fig. 1 (b1-b2)** and **Fig. 1 (c1-c2)**, the basal plane of ice is either slightly tilting up or down with respect to $V_p/G$ due to a minor orientation deviation between basal plane of ice and $V_p/G$. Tilting directions of two groups of ice tips of different tip shapes are consequently asymmetrical with respect to $V_p/G$. It should be noted that the ice morphology in **Fig. 1** is not the same to a case of bi-crystal competition of two crystal grains with different orientations [20, 21]. And two tilting directions of ice dendrites resulted from peculiar ice growth habit is exactly the morphology within a single ice crystal with orientation relationships of $\vec{V}_p \parallel \vec{G} \parallel \{0001\}, \vec{L} \parallel \{0001\}, \vec{V}_p(\vec{G}) \perp \vec{L}$.

### 3.2 Ice tip position as a function of time and determination of tilt angle $\alpha$

**Figure 2** illustrates an example of determination of tilt angle of ice tip based on data points $(X, Y)$ of a 0.1 M NaCl solution. Two-dimensional ice tip position $(X-Y)$ as a function of time $t$ was measured during in-situ experiments at time intervals $\Delta t$ of 5 or 10 seconds. It should be noted that two groups of ice tips were usually obtained in our experiments with one group tilting up and the other tilting down (eg. as shown in **Fig. 1 (c2)**). The following measurement can only be performed on one group of ice tips since movement of two groups of ice tips differ from each other. Ice tip position was first recorded in the in-situ movie (see a schematic photo from in-situ movie in **Fig. 2 (a)**) as a series of data points $(x, y)$ whose original point is chosen at the first data point $(x, y)$ of the ice tip being measured. By converting the $(x, y)$ in the moving frame of a corresponding pulling velocity of magnitude $V_p$ into $(X, Y)$ ($X = x + V_p \cdot \Delta t$, $Y = y$), the movement of ice tip with respect to the ground was obtained as a series of data points $(X, Y)$ as shown in **Fig. 2 (b)** under different pulling velocities. The linear regression of tip position data $(X, Y)$ of an ice dendrite growing under steady state yields its tilting direction of a given pulling velocity. Here tilt angle is defined as the difference between tilting directions of that at the lowest pulling velocity approximated as the preferred orientation of ice and a larger pulling velocity since when pulling velocity is low enough, the growth direction of ice tip is approximately determined by its preferred orientation ($\perp <0001>$). It is interesting to see in **Fig. 2 (b)** that some ice tips change its tilting direction with respect to the direction of $V_p/G$. Such a change of tilting direction with respect to thermal gradient is a challenging experimental evidence to a widely accepted explanation of tilted growth of ice platelets in freeze casting samples [28, 29] based on a combined effect of unparallel thermal gradient and preferred orientation of ice in unidirectional freezing. The ice tip switching from one side to another side of thermal gradient can not be explained in a conventional manner. And the physics of tilted growth of ice tips is essentially linked with its anisotropic nature.

### *3.3 Effect of thermal gradient, solute concentration and solute type on tilt angle $\alpha$*

Thermal gradient is an important parameter in unidirectional solidification. However, to our knowledge so far, it is hardly reported how thermal gradient affect tilt angle $\alpha$ of ice dendrite in unidirectional freezing of aqueous solutions. **Figure 3** shows the dependence of measured tilt angle $\alpha$ of ice dendrite on the applied thermal gradient $G$ in unidirectional freezing of a 0.1 M NaCl solution. **Figure 3** shows that within the investigated range of thermal gradient, $\alpha$ varies in a narrow range of no larger than 2.5°. The magnitude of thermal gradient does not have a significant effect on $\alpha$ of a 0.1 M NaCl solution.

**Figure 4** shows the effect of solute concentration and solute type on tilt angle $\alpha$ in unidirectional freezing experiments. In **Fig. 4**, both solute concentration and solute type can exsert a significant impact on $\alpha$. In details, for NaCl solutions (see the inset in **Fig. 4**), larger solute concentration seems to make $\alpha$ increase faster against the same pulling velocity, but the range of $\alpha$ is still as narrow as 2.5°. For PVA103/203 solutions, the effect of solute concentration is more obvious with broadened range from about 15° to about 45° due to increased concentration from 5 wt.% to 10 wt.%. It should be noted that the $\alpha$ as high as 45° in 10 wt.% PVA103 is close to the limiting value from free growth experiments in aqueous solution [15], possibly due to the significant influence of PVA103 macromolecules on kinetic anisotropy of ice growth via changes in several aspects, such as diffusion coefficient of water molecules in concentrated polymer solution [30], S/L interface attachment processes [31-33] and possible particle-induced surface instability [34]. Overall, in the experimental range tested, $\alpha$ in both **Fig. 3** and **Fig. 4** shows a trend of increment with pulling velocity, reaching a plateau or a limiting value which varies with both solute concentration and solute type. It is suggested that tilted growth of single ice crystal is sensitive to solute additives, which reveals that the unidirectional freezing morphology is intrinsically influenced by altered kinetic anisotropy of ice via solutes.

### *3.4 Tip undercooling for tilted growth of lamellar ice*

The tilted growth here is related to kinetic anisotropy of ice while the traditional explanation of tilted growth is from competition between solute diffusion and surface tension anisotropy [35]. The solute pileup around the tip can be revealed from tip undercooling of ice dendrites. Therefore, tip undercooling at which tilted growth of ice occurs is a traditional indicator of the tilted lamellar ice morphology. **Figure 5** shows tip undercooling for NaCl solutions is a monotonic decreasing function of pulling velocity, based on present experiments and also on previous experiments [36], and increased thermal gradient slightly increases tip undercooling of ice at roughly every pulling velocity. The decreased tip undercooling against pulling velocity indicates the reduced solute pile-up around the S/L interface. The tilt angle increases against increased growth velocity, indicating a changed interface kinetic anisotropy. However, the range of tilt angle hardly broadens compared with PVA103/203, possibly due to the limited influence of NaCl on the kinetic anisotropy of ice. The tip undercooling data of PVA solutions are also quite small [26], but the physical nature is far from being well understood compared to that in NaCl solution and no clear conclusion can be made on its relationship betwedn the tip undercooling and the tilted growth. Such low interface undercooling for tilted growth of ice also coincides with Levi's argument [23] which pointed out that both basal and edge planes of ice should enter a transition regime of crystal growth kinetics under some critical interface undercoolings lower than 1 $K$. The tilting of ice tip in aqueous solutions can occur at relatively low interface undercooling and the growth kinetics of ice is quite sensitive to growth parameters. Some transition in growth laws may occur under an interface undercooling lower than 1 $K$. Considering the complex interactions between solutes and ice growth, more precise unidirectional freezing experiments are needed to further quantify the possible variations of ice growth kinetics in different solutions.

### *3.5 Continuous change of tilting direction by abrupt stop of sample motion*

In order to further reveal the kinetic behavior of tilted growth of ice tip, we recorded morphology evolution of arrays of ice dendrites after an abrupt stop from steady state freezing in a 0.2 M NaCl solution and a 10 wt.% PVA203 solution after different time of intervals. In-situ movies for continuous change of tilting direction by abrupt stop of sample motion in a 0.2 M NaCl solution and a 10 wt.% PVA203 solution are provided in **Appendix B**. **Figure 6** vividly shows morphological evolution of ice in a 0.2 M NaCl solution (see **Fig. 6 (a-c)**) and also in a 10 wt.% PVA203 solution (see **Fig. 6 (d-f)**) from their steady state freezing under a given pulling velocity to an abrupt stop of sample motion. The results for NaCl solution (see **Fig. 6 (a-c)**) are not as obvious as in PVA203 solution (see **Fig. 6 (d-f)**), since as is mentioned in section 3.3, PVA103/203 solutions can produce much wider range of tilt angle than NaCl solutions in our experiments. In **Fig. 6 (d-f)**, tilt angle of ice tips gradually decreases as time goes on, forming bended lamellar morphology with decreased angle between lamellar ice tips tilting up and down. And the lamellar ice finally becomes almost parallel to the direction of thermal gradient. Such a phenomenon of continuous change of tilting direction in a single crystal is essentially a dynamic response of tilted growth of ice tips under a decaying kinetic undercooling to freeze. Since when a sample moving at a given pulling velocity comes to a halt on the freezing platform from an established thermal gradient at equilibrium with its surroundings, a new thermal equilibrium under static will gradually form, and the isotherm will move to the hot side of freezing platform and finally come to a halt, in which no kinetic undercooling is present in unidirectional growth of ice.

It should be noted that previous X-ray radiography experiments [37] also observed similar phenomenon at their initial instants of solidification, and relevant explanation was based on cooling conditions after instant nucleation of polycrystalline ice under an imposed temperature gradient. In the results of **Fig. 6**, however, ice orientation is fixed during the whole unidirectional freezing process and the tilted growth behavior is dominated by dynamic response of tilted lamellar ice tips to a decaying kinetic undercooling, which mainly embodies the sensitive nature of anisotropic ice growth. In previous investigations of unidirectional solidification in non-faceted systems [35], tilted growth usually refers to the case where there is a deviation between thermodynamically preferred orientation and thermal gradient, where the actual growth direction of dendritic tip is mainly controlled by solute diffusion effect at low growth velocities and by thermodynamically preferred orientation at large growth velocities. In such a traditional case, there will also be continuous change of growth direction of dendrites by abrupt stop of sample motion from steady state solidification to being static. However, the precondition for the traditional case is an angle between the preferred orientation of tip and the thermal gradient. And tilting of dendrites in traditional case can not simultaneously occur in two directions bilaterally with respect to thermal gradient. This differs from the case for tilted growth of ice in this paper, where two directions of tilting are simultaneously observed in single crystal ice. Besides, the variation of tilt angle in traditional case is constrained within the range between directions of thermal gradient and preferred orientation. In this paper, however, the preferred orientation of ice tip is set parallel to the thermal gradient, and the tilted growth is induced by its kinetic anisotropy which varies with imposed pulling velocity. Moreover, a continuous change covering the both sides of thermal gradient is observed as shown in the results of **Fig. 2 (b)**, which indicates a case distinct from traditional tilted growth of dendrites. And the sensitive growth property of ice reported here can be utilized to fabricate periodically varying porous microstructure in freeze casting materials by applying periodic growth conditions.

*3.6 Complex interactions between neighbouring ice tips*

During the experiments, some other interesting details concerning formation of side branches of ice dendrites are observed morphologically. In freeze casting, porous microstructure is closely related to ceramic/ice bridge and pore shape in samples [38]. Certainly, the possible interactions between neighbouring ice tips of sensitive nature can play a significant role in understanding the formation mechanism of freeze casting samples. In a 0.6 M NaCl solution, the brine volume or brine channel among ice cells/dendrites is much larger than NaCl solutions with lower concentrations and is therefore chosen for the following observation of such interactions. **Figure 7 (a-b)** shows how a side branch occurs on one side of an ice dendrite and grows in a tilted direction which is distinct from its neighbouring ice cells/dendrites. This side branch then grows faster than neighbouring ice cells/dendrites and is not overgrown by nerighbouring ice dendrites in a period of time as shown in **Fig. 7 (c)**. As time goes by, this side branch is later overgrown by a surrounding ice dendrite as shown in **Fig. 7 (d-e)**. Interestingly, another side branch occurs also on one side of this side branch and grows in another different direction, enabling the whole ice platelet to grow forward in a zigzag fashion as indicated by the black dash line in **Fig. 7 (f)**. Such process usually occurs periodically in some of the present experiments. Change of tilting direction in this interactive manner is unique and likely to be induced by thermal and/or solutal diffusion field of neighbouring ice cells/dendrites in combination with tilted growth behavior of ice. The possible interactions among tilted ice tips are previously discussed by Braslavsky [18] and Levi [23]. They pointed out that thermal interaction between two symmetrical ice tips with respect to the basal plane, though weak, can suppress the occurrence of side branch on inner side of two symmetrical ice cells/dendrites. But the interaction via solutal diffusion fields among neighbouring ice cells/dendrites, though hardly reported in previous investigations, seems to play a much significant role in these interactions. Further investigations are needed to explore the detailed physics behind it.

According to our observations, side branch instability [39] in freeze casting is speculated to operate in a similar way as in **Fig. 7 (a-c)**, which may be a direct result of interactions among neighbouring ice cells/dendrites in combination with the tilted growth habit of ice. It is reported that ice tips in both aqueous solution and colloidal suspensions usually adopt a peculiar dendritic morphology with side branches only on one side and the opposite side smooth [40]. In our experiments, similar dendritic ice tip morphology is also observed as shown in **Fig. 7**. Nevertheless, this tip morphology is not the only possible case in ice growth process, especially when undercooling is large enough. In some reports of ice growth from both liquid phase [14, 19, 23, 25] and vapor phase [41], side branches will occur on both sides of a primary ice dendrite and develop in directions almost parallel to their counterparts on the opposite side.

**4. Conclusion**

In conclusion, tilting behavior of lamellar ice tips in aqueous solutions were investigated by in-situ unidirectional freezing experiments. By in-situ measurement of ice tip movement, it is found that thermal gradient does not have a significant influence on tilted growth of ice within tested range of experiments. In contrast, solute type is more effective in changing the magnitude of tilt angle $\alpha$ of ice tips against pulling velocity. It is elucidated experimentally that tilted growth of ice tips is of sensitive nature and physically dominated by intrinsic growth habit of ice instead of a combined effect of unparalleled thermal gradient and preferred crystal orientation, which can be utilized to guide the fabrication of porous microstructure in freeze casting. Besides, tip undercooling of tilted growth of ice are measured to be mostly lower than 1 $K$, which coincides with the argument in previous investigations. Complex interactions among neighbouring lamellar ice via growth and overgrowing of side branch are also observed experimentally due to the combined effect of tilted growth of ice and temperature/solutal fields among arrays of lamellar ice.

In the future, it is important to carry out further investigations concerning tilted growth of ice both experimentally and theoretically through novel and accurate explorations since efforts on this problem is helpful to understanding the growth of ice crystal and other anisotropic materials.

**Acknowledgements**

The authors are grateful to Prof. K. G. Wang (Florida Institute of Technology) for his kind suggestions. This work was supported by the National Key R&D Program of China (Grant No.2018YFB1106003), National Natural Science Foundation of China (Grant No. 51701155), and the Fundamental Research Funds for the Central Universities (3102019ZD0402).

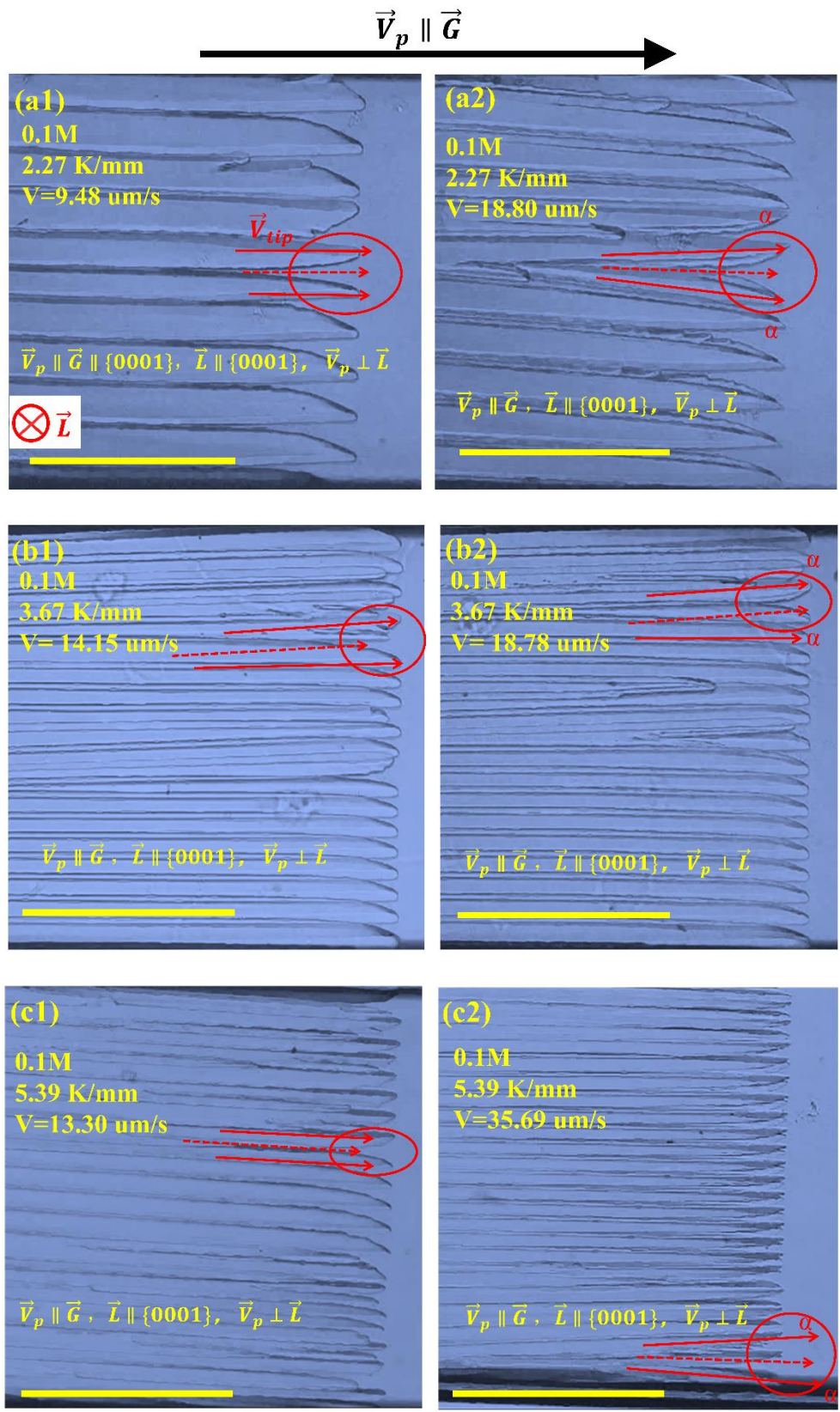

**Fig. 1** Typical tilted growth morphologies of single ice crystal in NaCl solutions under different pulling velocities and thermal gradients. Basal plane $\{0001\}$ is represented by a red dash arrow parallel to basal plane, and tip growth directions are represented by two red solid arrows. The black solid arrow on top of **Fig. 1** represents the directions of paralleled pulling velocity $\vec{V}_p$ and thermal gradient $\vec{G}$. $\vec{L}$ represents direction of incident light that is perpendicular to the observation plane. The crystal orientations of single crystal ice in **(a1-c2)** for directional freezing are manipulated to be slightly different, satisfying the relation of $\vec{V}_p \| \vec{G} \| \{0001\}, \vec{L} \| \{0001\}, \vec{V}_p(\vec{G}) \perp \vec{L}$ for **(a1-a2)**, $\vec{V}_p \| \vec{G}, \vec{L} \| \{0001\}, \vec{V}_p(\vec{G}) \perp \vec{L}$ for **(c-f)** in which the basal plane $\{0001\}$ is slightly **(b1-b2)** tilting up or **(c1-c2)** tilting down from $\vec{V}_p \| \vec{G}$. The scale bar in each figure was 250 um.

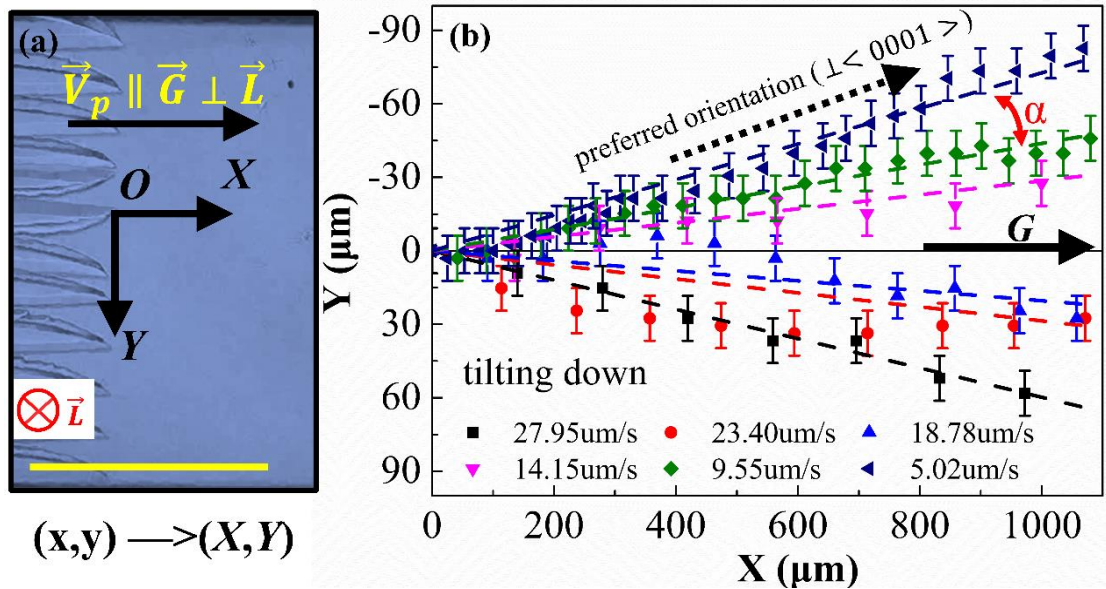

**Fig. 2** Determination of X-Y position of ice tip as a function of time in a 0.1 M NaCl solution. **(a)** A schematic photo showing how ice tip position was recorded in an in-situ movie. The original point is set as the tip position at the beginning of measurement. The pulling velocity $\vec{V}_p$ is set parallel to thermal gradient $\vec{G}$ and normal to the incident light $\vec{L}$ of microscope. **(b)** The actual tilting direction of an ice tip with respect to the thermal gradient $G$ can be evaluated by converting the slope of each linear fitting line of measured data points under each given pulling velocity (5.02 um/s, 9.55 um/s, 14,15 um/s, 18.78 um/s, 23.40 um/s and 27.95 um/s, respectively) into angle. Difference of tilting direction between the lowest pulling velocity approximated as parallel to the preferred orientation ($\perp<0001>$) of ice represented by a black dash arrow and another one pulling velocity of interest yields tilt angle $\alpha$ in the figure for an ice tip under this growth velocity. The thermal gradient is $G = 4.70$ K/mm. The time interval $\Delta t$ between each data point in this figure ranges from 5 seconds to 10 seconds.

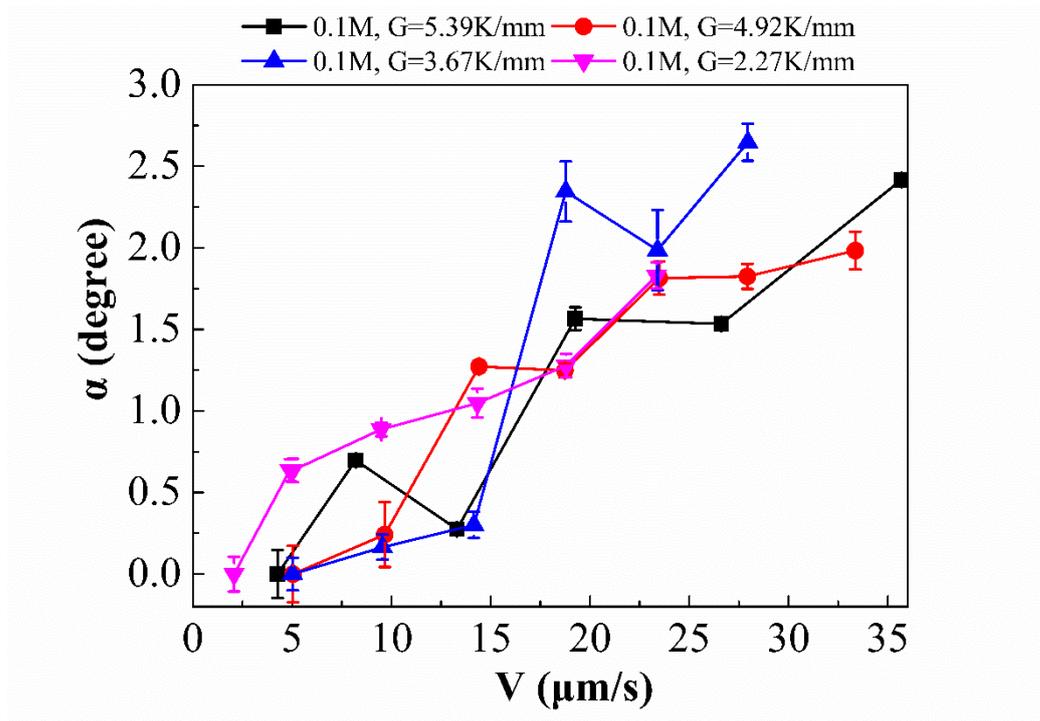

**Fig. 3** The effect of thermal gradient on $\alpha$-$V$ relation for ice growth in 0.1 M NaCl solution. The applied thermal gradients are $G = 2.27$ K/mm, 3.67 K/mm, 4.92 K/mm and 5.39 K/mm, respectively.

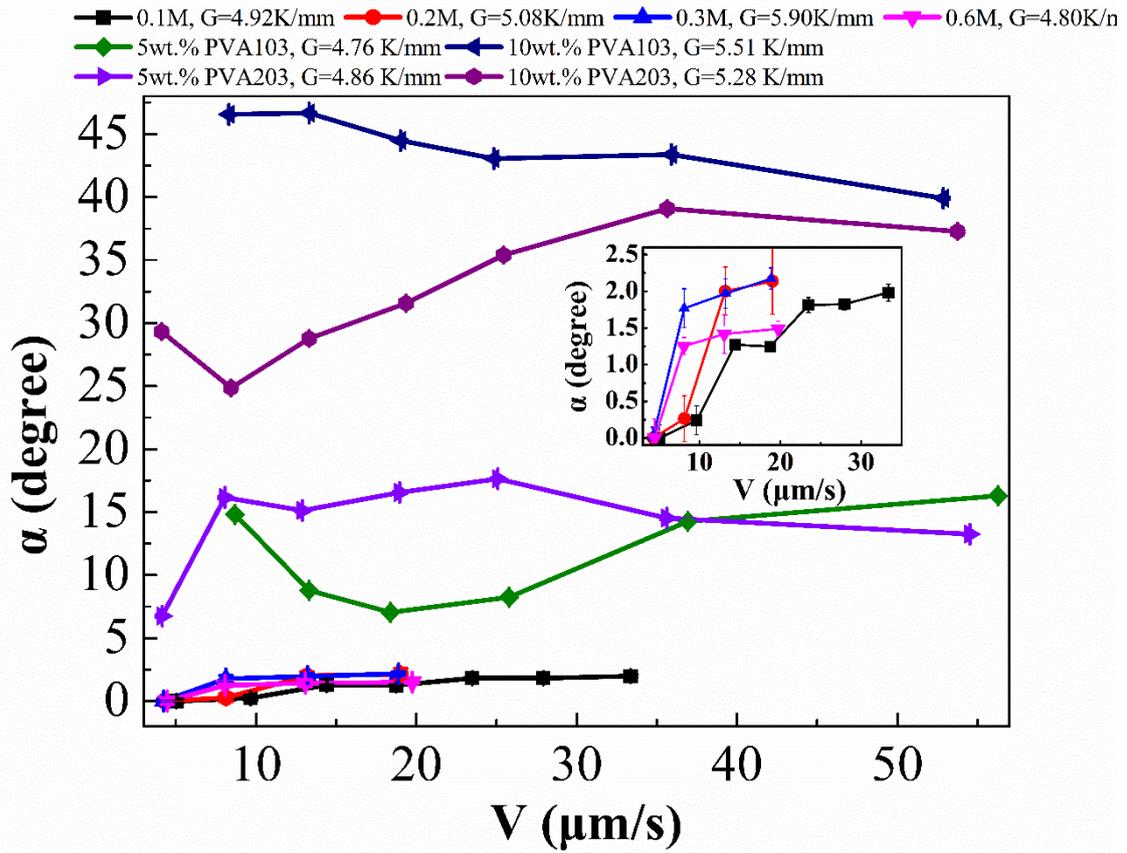

**Fig. 4** The effect of solute concentration and solute type on $\alpha$-$V$ relation for ice growth in NaCl solutions of varying concentrations (0.1M, 0.2M, 0.3M and 0.6M) and PVA203/103 solutions of varying concentrations (5 wt.% and 10 wt.%) and degrees of hydrolysis (86.5%-89% and 96.8%-97.6% hydrolyzed). The inset is a magnified figure of $\alpha$-$V$ relation for NaCl solutions. The applied thermal gradients for NaCl solutions are $G$ = 4.92 K/mm, 5.08 K/mm, 5.90 K/mm and 4.80 K/mm, respectively. The applied thermal gradients for PVA solutions are $G$ = 4.76 K/mm, 5.51 K/mm, 4.86 K/mm and 5.28 K/mm, respectively.

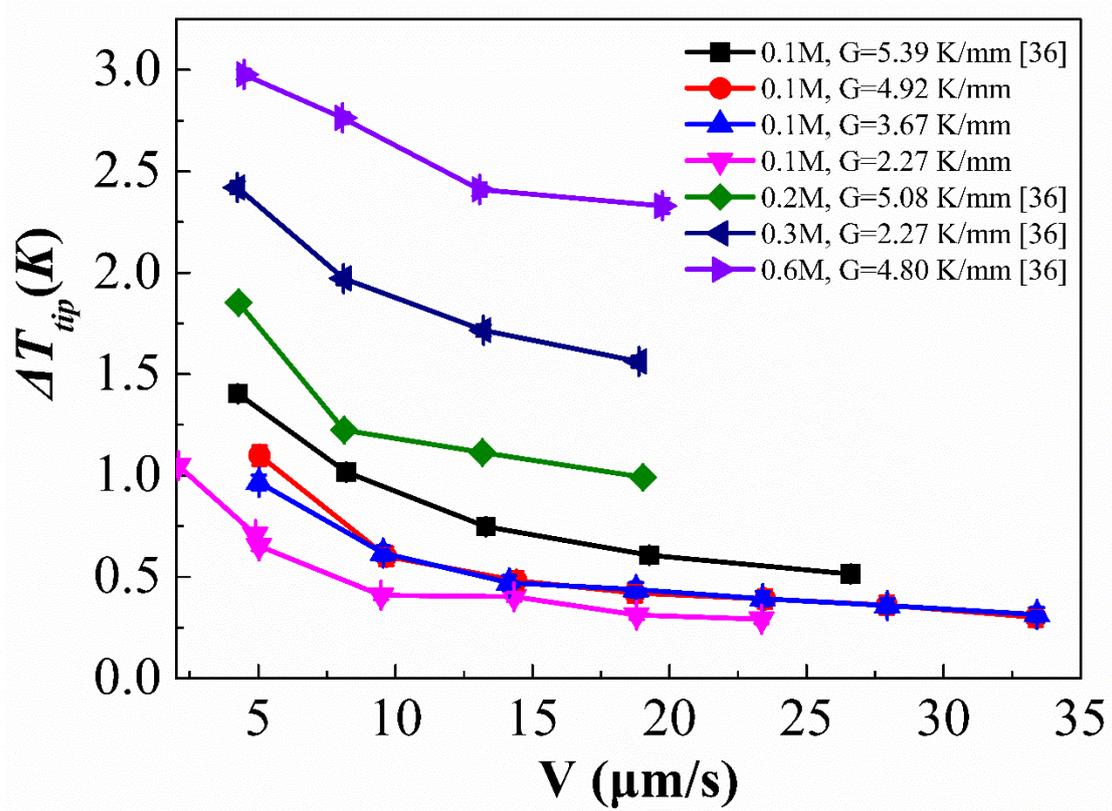

**Fig. 5** Variation of tip undercooling $\Delta T_{tip}$ of ice under steady state freezing as functions of pulling velocity, solute concentration and thermal gradient.

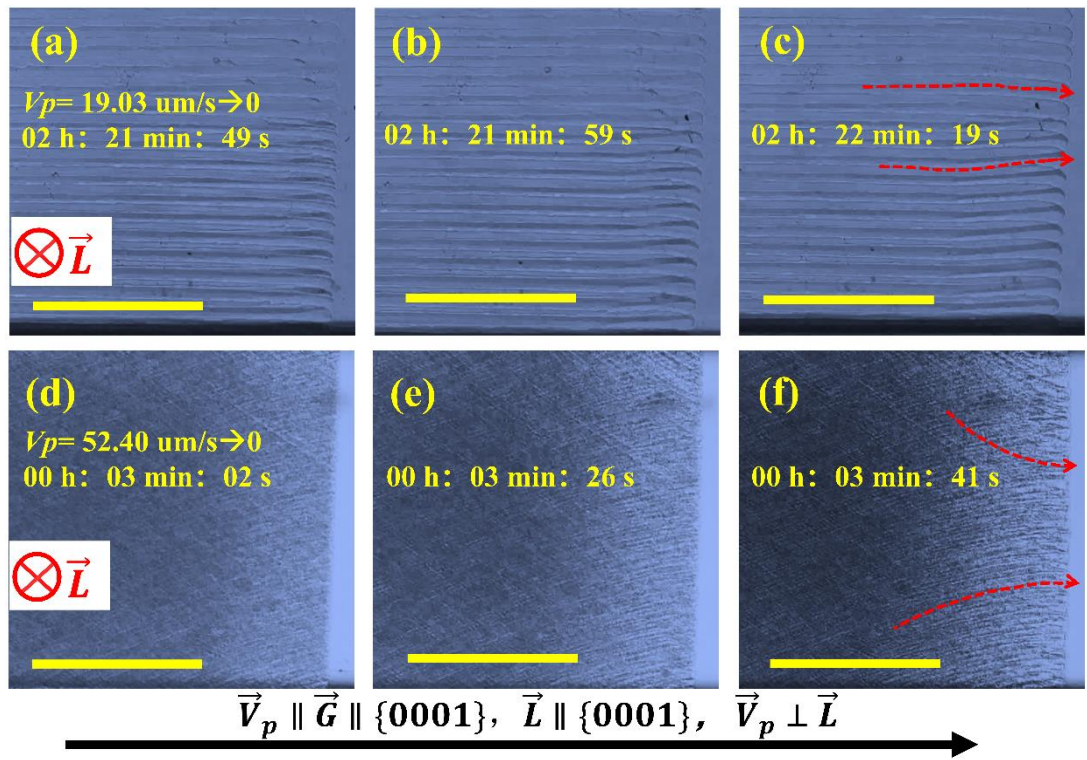

**Fig. 6** Morphology evolution of ice cells/dendrites in 0.2 M NaCl solution **(a-c)** and 10 wt.% PVA203 solution **(d-f)** from their steady state freezing to an abrupt stop of sample motion under a given pulling velocity. From **(a)** 02 hour: 21 min: 49 s, **(b)** 02 hour: 21 min: 59 s to **(c)** 02 hour: 22 min: 19 s, obvious change of tilting directions of ice cells/dendrites are observed in 0.2 M NaCl solution and are indicated in two red bended arrows in dash line in **(c)**. The initial steady-state conditions for **(a-c)** are $V_p$ = 19.03 um/s and $G$ = 5.08 K/mm. From **(d)** 00 hour: 03 min: 02 s, **(e)** 00 hour: 03 min: 26 s to **(f)** 00 hour: 03 min: 41 s, obvious change of tilting directions of ice cells/dendrites are observed in 10 wt.% PVA203 solution and are indicated in two red bended arrows in dash line in **(f)**. The initial steady-state conditions for **(d-f)** are $V_p$ = 52.40 um/s and $G$ = 5.29 K/mm. In these figures, the orientation relation of $\vec{V}_p \parallel \vec{G} \parallel \{0001\}, \vec{L} \parallel \{0001\}, \vec{V}_p(\vec{G}) \perp \vec{L}$ is satisfied as indicated by the black solid arrow below.

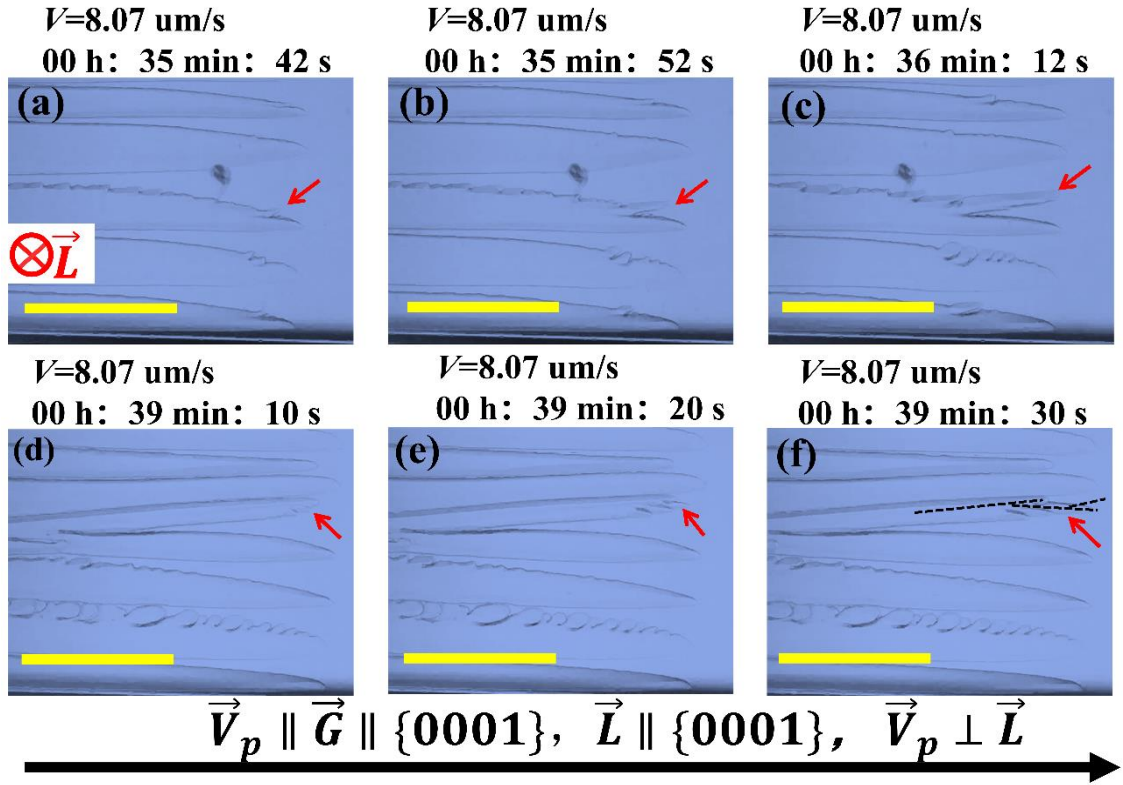

**Fig. 7** Interactions among ice cells/dendrites via **(a-c)** side branch and the subsequent **(d-f)** overgrowing between a secondary branch (indicated by red solid arrow) and a primary ice cell/dendrite above it in a 0.6 M NaCl solution. From **(a)** 00 hour: 39 min: 10 s, **(b)** 00 hour: 39 min: 20 s to **(c)** 00 hour: 39 min: 30 s, a side branch occurs and develops on upper side of an ice dendrite, which is indicated by the red solid arrow in **(a-c)**. From **(d)** 00 hour: 35 min: 42 s, **(e)** 00 hour: 35 min: 52 s to **(f)** 00 hour: 36 min: 12 s, the side branch can no longer grow upward and is overgrown by a neighbouring ice dendrite and another side branch occurs on bottom side of this side branch in **(a-c)** and the tip position of the newly occurred side branch gradually become parallel to its primary ice dendrite in **(a)** as is indicated by the red solid arrow in **(d-f)**. In **(f)**, the whole ice platelet indicated by the red solid arrow to grow forward in a zigzag fashion as indicated by the black dash line. The steady-state conditions for **(a-f)** are $V_p$ = 8.07 um/s and $G$ = 4.80 K/mm. In these figures, the orientation relation of $\vec{V}_p \parallel \vec{G} \parallel \{0001\}, \vec{L} \parallel \{0001\}, \vec{V}_p(\vec{G}) \perp \vec{L}$ is satisfied as indicated by the black solid arrow below.